\documentclass[sigconf]{acmart}

\copyrightyear{2022}
\acmYear{2022}
\setcopyright{rightsretained}
\acmConference[NLBSE'22]{The 1st Intl. Workshop on Natural Language-based Software Engineering}{May 21, 2022}{Pittsburgh, PA, USA}
\acmBooktitle{The 1st Intl. Workshop on Natural Language-based Software Engineering (NLBSE'22), May 21, 2022, Pittsburgh, PA, USA}
\acmDOI{10.1145/3528588.3528662}
\acmISBN{978-1-4503-9343-0/22/05}

\AtBeginDocument{%
  \providecommand\BibTeX{{%
    \normalfont B\kern-0.5em{\scshape i\kern-0.25em b}\kern-0.8em\TeX}}}

\usepackage{graphicx}
\usepackage{caption}
\usepackage{amsfonts}
\usepackage{amsmath}
\usepackage{booktabs}
\usepackage{siunitx}
\usepackage{hyperref}
\usepackage{subfigure}

\begin{document}
\title{CatIss: An Intelligent Tool for Categorizing Issues Reports using Transformers}

\author{Maliheh Izadi}
\email{m.izadi@tudelft.nl}
\affiliation{ 
\institution{Delft University of Technology}
\country{Delft, Netherlands}
}
\renewcommand{\shortauthors}{Maliheh Izadi}

\begin{abstract}
Users use Issue Tracking Systems
to keep track and manage issue reports in their repositories.
An issue is a rich source of software information
that contains different reports including a problem,
a request for new features, 
or merely a question about the software product.
As the number of these issues increases, 
it becomes harder to manage them manually.
Thus, automatic approaches are proposed 
to help facilitate the management of issue reports.
This paper describes \emph{\textbf{CatIss}}, 
an automatic \textbf{Cat}egorizer of \textbf{Iss}ue reports
which is built upon the Transformer-based pre-trained RoBERTa model. 
CatIss classifies issue reports into three main categories of 
\textit{Bug report}, \textit{Enhancement/feature request}, and \textit{Question}.
First, the datasets provided for the NLBSE tool competition 
are cleaned and preprocessed.
Then, the pre-trained RoBERTa model 
is fine-tuned on the preprocessed dataset.
Evaluating CatIss on about 
$80$ thousand issue reports from GitHub,
indicates that it performs very well 
surpassing the competition baseline,
TicketTagger, 
and achieving $87.2\%$ F1-score (micro average).
Additionally, as CatIss is trained on a wide set of repositories,
it is a generic prediction model, 
hence applicable for any unseen software project 
or projects with little historical data.
Scripts for cleaning the datasets, 
training CatIss 
and evaluating the model 
are publicly available.~\footnote{\url{https://github.com/MalihehIzadi/catiss}}

\end{abstract}


\maketitle

\section{Introduction}
Issue reports are used for communication, 
decision making, 
and collecting users' feedback in software repositories.
Any GitHub user can contribute 
to the progress of a project using issue reports.
Users open issue reports 
for different reasons including 
reporting a bug, 
requesting a new feature, 
asking for improvement,  
asking questions, 
and asking for support.

Issues in software repositories 
must have a \textit{title}, 
a \textit{description}, 
and a \textit{state} (open/closed).
They can also have additional data such as \textit{labels}, 
\textit{assignees}, \textit{milestone}, \textit{timestamps}, \textit{author association}, \textit{comments}, and more.
Issue description usually includes useful information 
about the reported problem 
and even code snippets to elaborate on the reported problem.
Moreover, each issue can have several labels 
such as \texttt{bug report} to denote the goal behind opening the issue.
Labels, as a sort of metadata, 
describe the goal and content of an issue.
They are mainly used for categorizing, managing, searching, and retrieving issues.
Thus, assigning labels to issues
facilitates task assignment, 
maintenance, 
and management of a software project.
Cabot et al.~\cite{cabot2015exploring} 
analyzed about three million non-forked GitHub repositories  
to investigate the label usage and its impact on resolving issues. 
They showed only about $3\%$ of these repositories had labeled issues.
Furthermore, in the repositories which incorporated issue labeling, 
only about $58\%$ of issues were labeled.
The authors showed addressing an issue
and the engagement rate 
both have a high correlation with the number of labeled issues in a repository~\cite{cabot2015exploring}.
Recently, Liao et al.~\cite{liao2018exploring} 
investigated the effect of labeling issues on issue management
and found labeled issues were addressed immediately, 
while unlabeled issues could remain open for a long time.

As the number of users and reported issues increases,
efficient and timely management of issue reports becomes harder. 
Team members should address these issues as soon as possible 
to keep their audience engaged 
and improve their software product.
Thus, automatic approaches for managing issues are proposed.
However, their accuracy can be improved by using contextual information better.

In this tool paper, 
using data-driven approaches,
The author fine-tunes a Transformer-based pre-trained RoBERTa-based model 
to predict the category of issue reports.
These categories are the three frequent reasons 
for opening an issue, namely 
\textit{Bug} reports, \textit{Enhancement} requests, and \textit{Questions}.
CatIss achieves $87.2\%$ F1-score 
as the micro average score for all three categories of issues.
Furthermore, CatIss obtains 89.6\%, 87.9\%, and 69.1\% of per class F1-score 
for bugs, enhancements, and questions, respectively.
As CatIss is built using a Transformer architecture,
it is able to better utilize the contextual information in issue reports,
hence providing more accurate predictions.
CatIss surpasses TicketTagger~\cite{ticket-tagger-scp,kallis2019ticket}, 
the baseline approach in all categories and for all metrics.
More importantly, based on the Recall score of classes,
CatIss significantly outperforms TicketTagger 
by $89\%$ for the \texttt{Question} category. 
Compared to Bug and Enhancement, 
``Question'' is the least-represented class in the datasets,
and consequently suffered from weak accuracy scores in previous studies.
However, CatIss is able to obtain much higher results.
CatIss is open source and can be accessed on GitHub.~\footnote{\url{https://github.com/MalihehIzadi/catiss}}
The Jupyter notebooks describes 
data processing, model training, and evaluation.
Note that this work is based on our previous study 
on the management of issue reports 
through categorization and prioritization~\cite{izadi2022predicting}.

\section{Tool Description}\label{sec:approach}
\begin{figure}[tb]
    \centering
    \includegraphics[width=\linewidth]{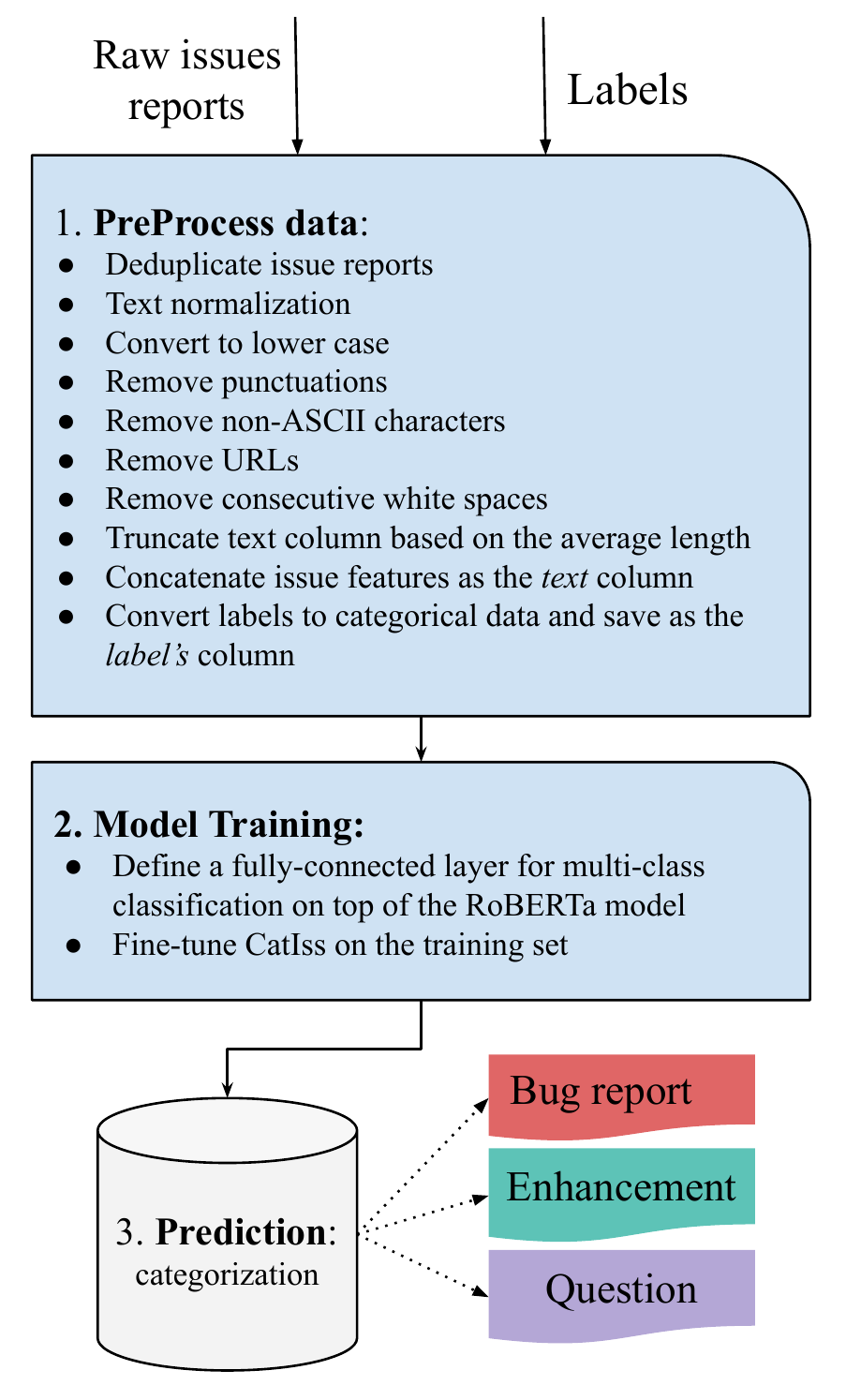}
    \caption{CatIss workflow}
    \label{fig:workflow}
\end{figure}

The author first cleans and pre-processes the dataset 
provided for the NLBSE tool competition~\cite{nlbse2022}.
Then, she proceeds to fine-tune 
the pre-trained Transformer-based RoBERTa model~\cite{liu2019roberta} 
on the preprocessed data, 
and finally, evaluates it on the competition test set.
In the following, CatIss will be described in more detail.
Figure~\ref{fig:workflow} 
presents a concise summary of the proposed approach.

\subsection{Preprocessing Data}\label{sec:preprocess}
Data processing techniques 
used to clean the provided training set are as follows.
Using the ``describe'' function of the Pandas dataframes,
it is evident that the training set consists of duplicate issue reports.
Thus, the author first removes the duplicate issues 
using their issue URL as their unique ID.
This process removes $26220$ 
from the $722899$ available issues in the training set.
Note that deduplication is only performed on the training set 
to avoid modifying the samples' distribution in the test set.

Next, the author perform normalization 
on the text of issue titles and bodies.
Normalization refers to the technique of 
replacing the content of a concept 
in the text with the name of that concept.
For instance, function names are extracted 
using regular expressions and replaced 
with a fixed token $\langle FUNCTION \rangle$. 
Text normalization is conducted with two objectives; 
(1) to keep the existence of a function in text, 
and (2) to reduce the vocabulary size
for the machine learning models~\cite{izadi2021topic}.

Then, punctuation marks, and non-ASCII characters 
are removed from both title and body of issues.
Next, the text in titles and bodies 
is converted to lower case.
The base URL of repositories (https://api.github.com/repos/)
is also removed from the \texttt{repository\_url} column as it does not provide additional information for the model.

After removing white spaces from the title and body columns,
the five remaining columns are concatenated together 
to construct the input for the model (excluding \texttt{issue\_url} column).
These five columns are:
(1) \texttt{issue\_create\_at}: the time that an issue was created,
(2) \texttt{issue\_author\_association}: the issue author role, 
(3) \texttt{repository\_url}: the repository `owner + repository' names,
(4) \texttt{issue\_title}, and finally
(5) \texttt{issue\_body}.

In the end, the \texttt{text} columns 
in both train and test datasets are truncated
based on the average and median number of tokens 
in issue titles and bodies of the \textit{train} set.
Note that one must not rely on 
the test set's statistical information 
as the model will not see the test set before training.
The average numbers of tokens for issue titles and bodies 
after cleaning are $6$ and $49$ tokens.
Also, the median number of tokens for the latter is about $100$ token.
Hence, the cutoff point is to $200$ tokens in the text column.
Note that the issue classes in the \texttt{label} column 
(bug, enhancement, and question)
are also converted to their categorical codes (0, 1, 2) 
and then fed to the model.

\subsection{Model Training}
\begin{figure}
    \centering
    \includegraphics[width=\linewidth]{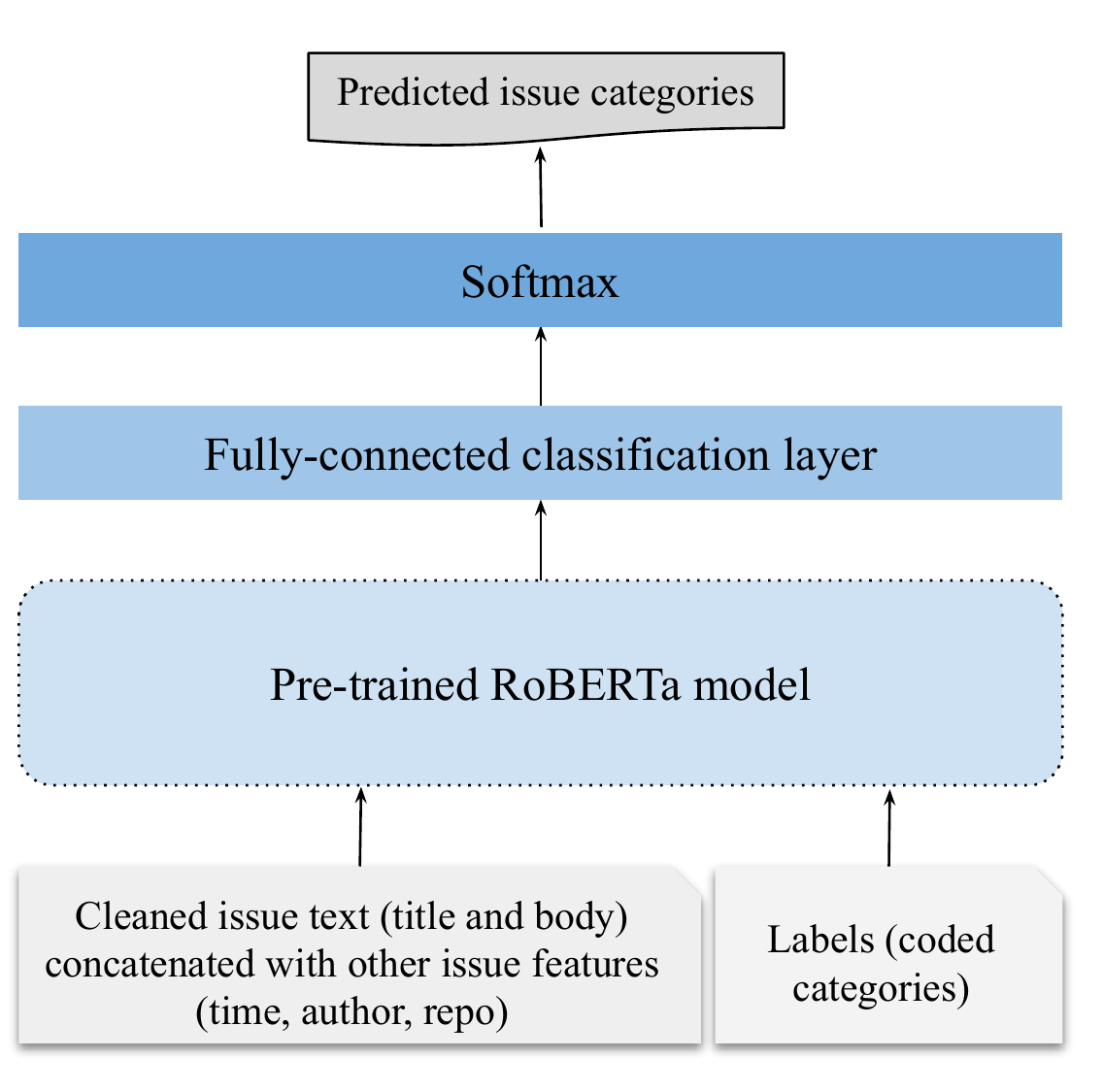}
    \caption{Fine-tuning RoBERTa for multi-class classification on issue reports}
    \label{fig:fine-tune}
\end{figure}

For the past few years, Transformers 
have significantly impacted the Natural Language Processing (NLP) domain.
Now, Pre-trained models such as Bidirectional Encoder Representations from Transformers (BERT) can be fine-tuned 
using additional outputs layers to create state-of-the-art models 
for a wide range of NLP tasks, 
without major task-specific architecture modifications. 
Based on our recent study on issue report management~\cite{izadi2022predicting}, 
CatIss adapts a Robustly-optimized BERT approach (RoBERTa) 
and fine-tunes it~\cite{liu2019roberta}.
RoBERTa, developed by Facebook~\cite{liu2019roberta}, 
includes additional pre-training improvements 
using only unlabeled text from the web, 
with minimal fine-tuning and no data augmentation. 
The authors modified the Masked Language Modeling (MLM) task
in the BERT using dynamic masking. 
The authors also eliminated the Next Sentence Prediction (NSP) task
since Facebook's analysis indicated 
that it actually hurts the model's performance. 
Finally, RoBERTa was trained using larger mini-batch sizes compared to BERT.
The author defines a multi-class classification layer 
on top of this model as the downstream task
and fine-tunes it for four epochs on the pre-processed training set.
Figure~\ref{fig:fine-tune} depicts the architecture of CatIss 
and the fine-tuning process.

\section{Experiment Setup}
In this section,
first datasets and their characteristics are reviewed, 
then the evaluation metrics are defined.
And finally, implementation details are presented.

\subsection{Datasets}
The training and test sets provided by the NLBSE workshop~\cite{nlbse2022}
are used for training and evaluating the model.
Training and test sets contain $722899$ and $80518$ issue reports, respectively.
The datasets contain various types of information on issue reports 
including time of creation, repository URL, issue URL, 
author association role, label, title, and body.
Issue reports are collected from $3093$ repositories hosted on GitHub.
There are six types of author association rules available in the dataset, namely 
OWNER, MAINTAINER, CONTRIBUTOR, COLLABORATOR, MEMBER, and MANNEQUIN.
Issues belong to one category of the three labels: 
bug, enhancement, and question.
About $8K$ of issues do not have a body text, 
that is, they are lacking any description.

\subsection{Metrics}
CatIss is evaluated using standard measures of evaluating classifiers, 
namely \textit{precision}, \textit{recall}, and \textit{F1-score}.
Equations~\ref{eq:precision}, \ref{eq:recall},
and \ref{eq:f1} compute the above measures 
for each class of issues, namely bug, enhancement and question.
Micro averages of the above measures are also reported.
TP, FP, TN, and FN indicates number of 
True Positives, False Positives, True Negatives, and False negatives, respectively.
\begin{equation}
\label{eq:precision}
P_c = \frac{TP_c}{TP_c + FP_c}
\end{equation}
\begin{equation}
\label{eq:recall}
R_c = \frac{TP_c}{TP_c + FN_c}
\end{equation}
\begin{equation}
\label{eq:f1}
F_{1,c} = \frac{2 \cdot P_c \cdot R_c}{P_c + R_c}
\end{equation}
\begin{equation}
\label{eq:pm}
P = \frac{\sum{TP_c}}{\sum{(TP_c + FP_c)} }
\end{equation}
\begin{equation}
\label{eq:rm}
R = \frac{\sum{TP_c}}{\sum{(TP_c + FN_c)} }
\end{equation}
\begin{equation}
\label{eq:f1m}
F_1 = \frac{2 \cdot P\cdot R}{P + R}
\end{equation}

\subsection{Implementation}
The author uses HuggingFace Transformers~\footnote{\url{https://huggingface.co}}, PyTorch~\footnote{\url{https://pytorch.org/}},
and SimpleTransformers~\footnote{\url{https://github.com/ThilinaRajapakse/simpletransformers}} libraries to train CatIss.
She fine-tunes the pre-trained RoBERTa model 
accessible through the HuggingFace API~\footnote{\url{https://huggingface.co/api/models/roberta-base}}.
The author also sets the learning rate to $3e-5$, 
the number of epochs to $4$, 
the maximum input length to $200$ 
and the training and evaluation batch sizes to $100$.
Experiments are conducted on a machine
equipped with  Ubuntu 16.04, 64-bit as the operating system,
two GeForce RTX 2080 GPU cards,
AMD Ryzen Threadripper 1920X CPU with $12$ core processors, 
and $64$G RAM.
Training lasts for four hours and $20$ minutes.

\section{Results}
Table~\ref{tab:results} presents CatIss's results 
on classifying issues of the test set
compared to the provided baseline, TicketTagger~\cite{kallis2019ticket}
and an additional baseline, Logistic Regression (LR) as a classical classifier.
As depicted, CatIss is outperforming the baselines, 
and achieving an F1-score of $0.872$ (micro average).
Note that the more populated categories 
(\textit{Bug}, and \textit{Enhancement}), 
obtain higher classification results for all metrics 
compared to the less-represented class \textit{Question}.
This is probably due to two reasons; 
(1) the fewer number of samples for this class,
(2) the diversity of information and vocabulary in this category.
Nonetheless, CatIss significantly outperforms TicketTagger 
for the \textit{Question} category (by $89\%$ based on the Recall). 
Compared to Bug and Enhancement, 
``Question'' is the least-represented class in the datasets,
and consequently has suffered from weak accuracy scores 
in the previous studies.
However, CatIss, utilizing contextual information 
using its transformer architecture 
and the knowledge transferred from the pre-trained model, 
is able to obtain higher scores for this class, 
specifically for Recall and F1-measure.
It is worth mentioning that the author 
also performed a data ablation study.
The results of this experiment indicated that including 
and concatenating the five sources of information reviewed 
in Section~\ref{sec:preprocess} provide better results.
\begin{table*}[tb]
\caption{Classification Results}
\centering
\begin{tabular}{cccc|ccc|ccc}\toprule
Metric      
& \multicolumn{3}{c}{CatIss} 
& \multicolumn{3}{c}{TicketTagger~\cite{kallis2019ticket}} 
& \multicolumn{3}{c}{Logistic Regression}\\
\cmidrule{2-10}
            & P & R & F1 & P & R & F1 & P & R & F1\\\midrule
Bug         
& \textbf{0.894} & \textbf{0.897} & \textbf{0.896}  
& 0.831 & 0.872 & 0.851 
& 0.841 & 0.867 & 0.854 \\\midrule
Enhancement 
& \textbf{0.874} & \textbf{0.885} 
& \textbf{0.879}  & 0.815 & 0.846 & 0.831 
&0.822 & 0.850 & 0.835 \\\midrule
Question    
& \textbf{0.720} & \textbf{0.664} & \textbf{0.691} 
& 0.652 & 0.350 & 0.455 
&0.655 & 0.432 & 0.521 \\\midrule
Micro Avg   
& \textbf{0.872} & \textbf{0.872} & \textbf{0.872}  
& 0.816 & 0.816 & 0.816 
& 0.822 & 0.822 & 0.822 \\
\bottomrule
\label{tab:results}
\end{tabular}
\end{table*}

\section{Related Work}\label{sec:related}
Kallis et al.~\cite{kallis2019ticket,ticket-tagger-scp} has proposed \textit{TicketTagger}, 
a tool based on FastText for classifying issues to three categories of 
\texttt{Bug}, \texttt{Enhancement} and \texttt{Question}.
These categories are among the default labels of the GitHub issue system.
They trained their model on the text (title and description)
of $30K$ issue reports from about $12K$ GitHub repositories.
Their evaluation reports $82\%$, $76\%$, and $78\%$ of precision/recall scores 
for three classes of Bug, Enhancement, and Question, respectively.
Recently, BEE was proposed by Song and Chaparro~\cite{song2020bee}, 
which uses the pre-trained model of TicketTagger to label issues. 
Then it proceeds to identify the structure of bug descriptions 
from predicted reports that are predicted to be a bug 
in the issue-objective prediction phase.
In our recent study~\cite{izadi2022predicting},
we propose two models to first extract issues' objectives 
and then prioritize them using the predicted objectives and other issue features.
CatIss is based on this study.

\section{Conclusions}\label{sec:conclude}
CatIss is an automatic categorizer of issue reports
that is built upon the Transformer-based pre-trained RoBERTa model. 
CatIss categorizes issue reports into three main categories of 
\textit{Bug}, \textit{Enhancement}, and \textit{Question}.
CatIss is trained on the provided training set
for the NLBSE tool competition 
and achieves $87.2\%$ F1-score (micro average) on the test set.
Specifically, CatIss significantly 
outperforms TicketTagger, the main baseline
by $89\%$ based on the Recall score of the Question class.
Scripts for cleaning the datasets, 
training CatIss 
and evaluating the model 
are publicly available.~\footnote{\url{https://github.com/MalihehIzadi/catiss}}

\bibliographystyle{ACM-Reference-Format}
\bibliography{main}
\end{document}